\documentclass[11pt, a4paper, logo, onecolumn, nonumbering]{deepmind}

\usepackage[authoryear, sort&compress, round]{natbib}
\usepackage{subcaption}
\usepackage{amsmath}
\usepackage[utf8]{inputenc} 
\usepackage[T1]{fontenc}    
\usepackage{url}            
\usepackage{booktabs}       
\usepackage{amsfonts}       
\usepackage{nicefrac}       
\usepackage{microtype}      
\usepackage{xcolor}         
\usepackage{subcaption}
\usepackage{amsmath}

\newcommand{\R}{\mathbb{R}}

\usepackage{dsfont}
\newcommand{\ind}{{\mathds{1}}}

\title{A learning agent that acquires social norms from public sanctions in decentralized multi-agent settings}

\correspondingauthor{vinitsky.eugene@gmail.com, jzl@deepmind.com}

\reportnumber{} 

\renewcommand{\today}{September 2022}

\author[1, 2]{Eugene Vinitsky}
\author[1]{Raphael K{\"o}ster}
\author[1]{John P. Agapiou}
\author[1]{Edgar Du{\'e}{\~n}ez-Guzm{\'a}n}
\author[1]{\\Alexander Sasha Vezhnevets}
\author[1]{Joel Z. Leibo}

\affil[1]{DeepMind}
\affil[2]{UC Berkeley}

\begin{abstract}
Society is characterized by the presence of a variety of social norms: collective patterns of sanctioning that can prevent miscoordination and free-riding. Inspired by this, we aim to construct learning dynamics where potentially beneficial social norms can emerge. Since social norms are underpinned by sanctioning, we introduce a training regime where agents can access all sanctioning events but learning is otherwise decentralized. This setting is technologically interesting because sanctioning events may be the only available public signal in decentralized multi-agent systems where reward or policy-sharing is infeasible or undesirable. To achieve collective action in this setting we construct an agent architecture containing a classifier module that categorizes observed behaviors as approved or disapproved, and a motivation to punish in accord with the group. We show that social norms emerge in multi-agent systems containing this agent and investigate the conditions under which this helps them achieve socially beneficial outcomes.
\end{abstract}

\newcommand{\BibTeX}{\rm B\kern-.05em{\sc i\kern-.025em b}\kern-.08em\TeX}

\usepackage{dsfont}

\begin{document}

\maketitle

\section{Introduction}

Autonomously operating learning agents are becoming more common and this trend is likely to continue accelerating for a variety of reasons. First, cheap sensors, actuators, and high-speed wireless internet have drastically lowered the barrier to deploy an autonomous system. Second, autonomy creates the possibility of learning ``on device'', keeping experience local and off of any central servers. This makes it easier to comply with privacy requirements~\citep{kairouz2019advances} and increases robustness by removing a single point of failure. Third, the autonomous approach is a potentially better fit for never-ending life-long learning~\citep{platanios2019jelly} since it does not require periodic syncing with updated centralized models. Indeed fully autonomous agents do not require any train-test separation at all, a property thought to be important for establishing open-ended autocurricula~\citep{leibo2019autocurricula, stanley2019open}.

However, the presence of multiple interacting autonomous systems raises a host of new challenges. Autonomously operating learning agents must be robust to the presence of other learning agents in their environment (e.g.~\cite{crandall2018cooperating, carroll2019utility}). A significant issue that arises in the case of autonomous and decentralized learning agents is how to align their incentives. Working together is often difficult when agents all may prefer to maximize their own rewards at one another's expense. For instance, autonomous vehicles from multiple competing technology companies must share the road with one another and with human drivers (e.g.~\cite{liang2019federated}). Each car (company) ``wants'' to ``selfishly'' transport its users as quickly as possible. However, road congestion emerging from poor coordination negatively affects everyone. Human users also participate in these multi-agent systems, with even more autonomy. For instance, city neighborhoods compete with each other to reshape their roadways to incentivize driving apps to route traffic to other neighborhoods~\citep{ccolak2016understanding}. Fundamentally, in collective action problems, letting agents egoistically optimize their own reward leads to a worse outcome for everyone than if all cooperate. This problem is particularly difficult if different ways to cooperate exist and agents have divergent preferences over the outcomes. In this case, \emph{uncoordinated} cooperation may be no better than mutual defection. In these cases it is difficult for a consensus to emerge.  

To address such social dilemmas we take inspiration from a mechanism that human societies use to resolve some of the collective action problems they face: \emph{social norms}---group behavior patterns that are underpinned by decentralized social sanctioning (approval and disapproval: equivalently, reward and punishment)~\citep{fehr2004social, wiessner2005norm, balafoutas2014direct}. Social norms enable cooperative behavior in a wide variety of collective action problems which otherwise would fail due to free-riding and defection. Human civilization is thick with social norms~\citep{tomasello2013origins, young2015evolution, henrich2021origins}. They are critical to our welfare because they discourage harmful behaviors (e.g.~smoking in public places) and encourage beneficial behaviors (e.g.~charitable donation and voting)~\citep{bicchieri2016norms, nyborg2016social}. Social norms are also important components in institutional solutions to small community scale natural resource management problems~\citep{ostrom2009understanding, hadfield2013law} and aid large-scale collective actions like labor negotiations and democratic elections~\citep{olson1965logic, granovetter1978threshold, marwell1993critical, ostrom1998behavioral}. 

The critical assumption that will enable our agents to learn social norms by decentralized multi-agent reinforcement learning is that of \emph{public sanctioning}. In this paradigm, there are discrete events when agent $i$ makes their disapproval of agent $j$ known, an event that is typically punishing to the recipient in the sense of reinforcement learning. These events are considered to be public so learning may be conditioned on knowledge of all sanctioning events from any agent to any other agent. This paradigm has several positive features. For instance, it allows for the possibility of human participants sanctioning autonomous machines through the same ``API'' that the machines use to sanction one another. For instance, human drivers and self-driving cars could honk at each other or leave 1-star reviews. As sanctions occur and are stored, databases of sanctioning events could enable agents to adapt to local customs like differing driving patterns between cities. 

We construct an agent architecture that can use public sanctions to spark the emergence of social norms in a multi-agent reinforcement learning system. Our approach, which we call \emph{Classifier Norm Model} (\textbf{CNM}), takes inspiration from some of the key features that give efficacy to human social norms. First, social norms divide behavior into approved and disapproved categories. That is, they are classifiers~\citep{hadfield2014microfoundations}. Each agent has its own private representation of the group's schema for what constitutes approved behavior. In our model, agents view other actors in the scene and generate a prediction for whether society at large would approve or disapprove of their behavior~\citep{boyd2021arbitration}. Second, we assume that both human and artificial agents are intrinsically motivated to disapprove of behaviors that their group disapproves of~\citep{fehr2004social, xiao2005emotion, boehm2012moral}. 

We show that \textbf{CNM} magnifies emergent joint activity patterns that arise by chance in early exploratory learning. This ``bandwagon'' effect simultaneously pushes agents to cooperate and encourages them to cooperate in the same way as one another. Thus it mitigates the two fundamental dilemmas within each collective action problem: the start-up and free rider problems (terminology from~\cite{marwell1993critical}). In two complex collective action problems, we show that groups of \textbf{CNM} agents acquire beneficial social norms that decrease free-riding and coordinate cooperative actions, thereby causing higher per-agent returns. Next, we consider our results in light of arbitrariness properties of real-world social norms. That is, specific norms are not always beneficial relative to counterfactual situations where other norms prevail (different ways of cooperating)~\citep{ostrom2009understanding, bicchieri2016norms}. This is a key property of real-world norms and our model also captures it. Finally, we analyze the \textbf{CNM} agent architecture with ablation experiments to understand which architectural assumptions are key to our results.

\section{Related work}

Significant progress in multi-agent reinforcement learning has occurred over the last few years driven by rapid innovation in a paradigm where researchers assume that even though policies must ultimately be executed in a decentralized manner (without communication at run time), they can be trained offline beforehand in a centralized fashion. This paradigm is called centralized training with decentralized execution (CTDE)~\citep{lowe2017multi, sunehag2018value, rashid2018qmix, foerster2018counterfactual, iqbal2019coordinated, baker2020emergent}. Many  algorithms in this class~\citep{lowe2017multi, baker2020emergent} take an actor-critic approach and employ a centralized critic that takes in observations from all agents to produce a single joint value. One algorithm called OPRE maintains the division between training and test phases but does not learn a centralized critic. Instead in OPRE each agent learns its own critic but all critics are conditioned on the observations of the other players. This is interpreted as information available in ``hindsight''~\citep{vezhnevets2020options}. Other techniques make extensive use of the centralized regime by expanding and pruning the support of policies in each rollout; this includes algorithms like PSRO~\citep{lanctot2017unified} and XDO~\citep{mcaleer2021xdo}.

A rather different class of models takes the approach of constraining the \emph{kind} of information that can be communicated between agents, instead of constraining the time (training time versus test time) of its communication. These models avoid the need for explicit training and testing phases. They can be executed online and maintain full decentralization except for the specific data they need to communicate. Some researchers have studied the case where no information at all is communicated between agents. However this approach cannot usually resolve social dilemmas or coordinate on beneficial equilibria when multiple equilibria exist unless special environmental circumstances prevail~\citep{leibo2017multiagent, perolat2017common, koster2020model}. A few algorithms eschew training/testing but still cannot be considered fully decentralized since they require each player to be able to access the policies of other players~\citep{jaques2019social, foerster2018learning}. Most algorithms in this class that can robustly find socially beneficial equilibria in collective action problems require public rewards~\citep{hughes2018inequity, peysakhovich2018prosocial, eccles2019learning, wang2019evolving, mckee2020social, gemp2020d3c} or the ability to redistribute rewards amongst agents'~\citep{lupu2020gifting, wang2021emergent}. This class of algorithms assumes that while they are learning all agents will have real-time access to one another's rewards. 

However, making reward data public is undesirable for several reasons. (A) Agent designers may want to alter reward functions without affecting the larger multi-agent system. (B) Agent designers may be prohibited from sharing their agents' reward function on privacy grounds, for instance, if they constructed it from individual user data~\citep{kairouz2019advances}, or their reward functions may be proprietary. (C) Humans may inhabit the same multi-agent system as artificial agents. This is most apparent in autonomous vehicle applications. Humans cannot publicize their instantaneous reward signals, but both human-driven and self-driven cars can honk their horn to admonish others for bad driving.

In the real world, social norms need not be beneficial. For example they may ossify inefficient economic systems or unfairly discriminate against classes of people~\citep{mackie1996ending, akerlof1976economics, bicchieri2016norms}. In other cases, social norms can be ``silly rules'' that are neither directly harmful nor helpful~\cite{hadfield2019legible, koster2022spurious}. Yet some social norms are clearly helpful, like those that discourage harmful behavior. There are two main mechanisms through which beneficial social norms function: (A) stabilizing cooperation in social dilemma situations as the sanctioning can transform the payoffs into a game with new equilibria~\citep{ullmann1977emergence, kelley2003atlas} and (B) equilibrium selection. Here the question is how it can be predicted which equilibrium a society will select, given that multiple equilibria exist for the social situation in question (e.g.~\cite{lewis1969convention}). In this case the norm is a piece of public knowledge on which individuals may condition their behavior to rationally coordinate their actions with one another~\citep{vanderschraaf1995convention, gintis2010social, hadfield2012law}. Naturally, these two functions are often intertwined (e.g.~\citep{bicchieri2006grammar}). In this spirit, social norms have been treated in AI research as equilibria of repeated normal form games~\citep{shoham1997emergence, sen2007emergence}.

Recent work has aimed to study social norms in more complex models of human societies. One line of research has represented social norms with classifiers that label a behavior's social approval or disapproval. For instance, \cite{boyd2021arbitration} studied how such a classifier can interact positively with a reputation-based account of cooperation in iterated matrix games and \cite{koster2022spurious} demonstrated the potential benefits of a ``hand-crafted'' (i.e. not learned) classifier on the learning dynamics of enforcement and compliance behavior in multi-agent reinforcement learning.

\section{Multi-agent reinforcement learning with sanctions}

The formal setting for multi-agent reinforcement learning with sanctions is an $N$-player partially observed general-sum Markov game (e.g.~\cite{shapley1953stochastic, Littman94markovgames}) augmented with a concept of sanctioning and a public observation function that indicates when a player has sanctioned another player and with what valence (approval or disapproval).

\subsection{Definition: Markov game}

At each state $s \in \mathcal{S}$ of a Markov game, each player $i \in I = \{1, \dots, N\}$ takes an action $a_i \in \mathcal{A}_i$. Players cannot perceive each state directly, but instead receive their own $d$-dimensional partial observation of the state $o_i \in \R^d$, which is determined by the observation function $\mathcal{O} : \mathcal{S} \times I \rightarrow \R^d$. After the players' joint action $\vec{a} = (a_1, \dots , a_N)$, the state changes according to the stochastic transition function  $\mathcal{T} : \mathcal{S} \times \mathcal{A}_1 \times \! \cdots \! \times \mathcal{A}_N \rightarrow \Delta(\mathcal{S})$, where $\Delta(\mathcal{S})$ denotes the set of discrete probability distributions over $\mathcal{S}$. After each transition, each player $i$ receives a reward $r_i \in \R$ according to the reward function $\mathcal{R} : \mathcal{S} \times \mathcal{A}_1 \times \! \cdots \! \times \mathcal{A}_N \times \mathcal{S} \times I \rightarrow \R$.

We extend this standard definition to include the additional concept of \emph{sanctioning}. Sanctioning is assumed to be something that one player does to another player (it is dyadic). All players are assumed to have common knowledge of which events are sanctioning events and their valence (whether they are approval or disapproval).

\subsection{Definition: Markov game with sanctions}
\label{sec:markov_game_with_sanctions}

We define a \emph{sanctioning opportunity} as a situation where one agent can sanction another agent by taking an action that causes them a reward or punishment. The reward implications may be indirect. Sanctioning may not produce any instantaneous reward. For instance, an action may be punishing if it causes its recipient's future rewards to be less probable or delayed. There may be many different ways for agents to cause each other reward and punishment. Not all actions that cause reward or punishment are sanctioning actions. The Markov game with sanctions model stipulates that certain specific events are sanctioning events. It assumes all the agents have common knowledge of which events are sanctioning events.

If agent $i$ has an opportunity to sanction agent $j$ and chooses to punish them with its next action we call this a \emph{disapproval event}. If agent $i$ has a sanctioning opportunity but does not choose to punish agent $j$ with its next action we call this an \emph{approval event}\footnote{Symmetrically, it is possible to define approval events to be when the agent with the opportunity takes an action to reward the other agent and disapproval events to be when it does not do so (positive sanctioning). However we do not consider that case here. We made this choice because the bulk of the literature on sanctioning and social norms is primarily concerned with negative sanctioning~\citep{baldwin1971power, sep-social-norms}.}. Sanctioning opportunities are often situations where agent $i$ and agent $j$ are physically near one another, but in general they need not be. For instance, a user of a decentralized restaurant recommendation platform may leave a 1-star review to show their disapproval of a restaurant they visited several days prior.

Formally, for any given state $s \in \mathcal{S}$, let the set of sanctioning opportunities be given by $\mathcal{J}(s) \subseteq I^2$, where $(i, j) \in \mathcal{J}(s)$ whenever agent $i$ has a sanctioning opportunity towards agent $j$. Note that $\mathcal{J}(s)$ may be empty if no agent has a sanctioning opportunity in state $s$, and at the other extreme $\mathcal{J}(s) = I^2$ when every agent can sanction every other agent (including themselves). 

In this work, agents show their disapproval by emitting a zapping beam that has a punishing effect on any agent hit by it. A sanctioning opportunity $(i, j)$ therefore exists only if agent $i$ is physically in range to zap agent $j$.

\subsection{Definition: Markov Game with public sanctions}
\label{sec:sanction_obs}

A Markov game with public sanctions is a Markov game with sanctions that has been additionally augmented with a \emph{sanctioning observation} that is shared by all players. At each state, in addition to their individual observation $o_i$, each player $i$ also receives a sanctioning observation $g \in \mathcal{G}$, defined by the sanction-observation function $\mathcal{B} : \mathcal{S} \rightarrow \mathcal{G}$. This observation broadcasts information on the occurrence of sanctioning to all players.

It is natural to regard the public sanctioning observation as arising from a process of gossip whereby knowledge of who transgressed rapidly diffuses through a community. This interpretation may be useful for research that applies the Markov game with public sanctions model to study social-behavioral phenomena. On the other hand, when we think of modern technology like autonomous vehicles through this lens then we usually envision the public sanctioning observation as a kind of database to which all cars may read and write.

Let $\mathcal{C}(s, i, j)$ be the \emph{context} of sanctioning opportunity $(i, j) \in \mathcal{J}(s)$---the perspective of the decision-making agent leading up to its choice to approve/disapprove. In general, $\mathcal{C}(s_t, i, j) = (o^{(i)}_{0:t}, a^{(i)}_{0:t-1})$, the full history of the decision-making agent's individual observations and actions; however, it is also possible to use less context. For instance, in the environments we study here, agents change color as a function of their recent behavior. Thus it is sufficient to choose $\mathcal{C}(s_t, i, j) = o^{(i)}_t$, the current observation of the agent with the sanctioning opportunity. E.g.~think about a child stealing a cookie. If when you encounter them they still have chocolate all over their face then you need not have directly observed their transgression to disapprove of their behavior. 

Finally, let $\mathcal{Z}(s, \vec{a}, i, j) \in \{0, 1\}$ be a binary indicator of whether the actions $\vec{a}$ taken in state $s$ resulted in a \emph{disapproval} event (of $j$ by $i$). In this work we define $\mathcal{Z}(s, \vec{a}, i, j) = 1$ if agent $i$ zaps agent $j$.

Putting everything together, we get a sanction-observation function that, at time $t$, returns a view of the sanctioning opportunities at time $t-1$, the sanctioning decisions made at those opportunities, and the context for those decisions:
\begin{align*}
\mathcal{B}(s_{t-1}, \vec{a}_{t-1}) = \{&(i, j, c, z) \text{ such that } \\
&(i, j) \in \mathcal{J}(s_{t-1}) \text{ and } \\
&c = \mathcal{C}(s_{t-1}, i, j) \text{ and } \\
&z = \mathcal{Z}(s_{t-1}, \vec{a}_{t-1}, i, j) ) \}
\end{align*}
Note that this depends on the previous state $s_{t-1}$ and actions taken $\vec{a}_{t-1}$, but it can still be represented as $\mathcal{B}(s_t)$ by augmenting the state to include prior observations or actions.

\subsection{Interpretation of the definitions}
\label{sec:interpretation_of_mg_with_public_sanctions}

To build intuition for what constitutes sanctioning, consider a human driving along the highway. We assume that humans dislike having a car horn honked at them. This attitude may only partly depend on the intrinsically aversive nature of the honking sound itself. Most of the negative experience of being honked at derives from understanding the sound's cultural context. Drivers honk when they want to admonish other drivers for their bad behavior. Thus being honked at may be aversive through a guilt mechanism (``I am sorry I transgressed'') or through an anger/reciprocity mechanism (``how dare you say I transgressed!''). No matter the cause, the important thing is common knowledge on the part of the whole driving community that honking is meant to be admonishing. 

Of course drivers do not always honk to sanction one another. For instance, they also honk to alert one another of danger. There is plenty of scope for disagreement concerning whether a given honk was intended as sanctioning or alerting. In this, sanctioning is no different from any other form of communication where ambiguity is pervasive but humans are nevertheless able to recover their partner's intent. In the case of honking it is usually obvious from context that a given honk was intended as sanctioning. Sometimes, if worried the current context may not make their meaning clear, individuals may seek to resolve ambiguity by adding an extra ``flourish'' to their honk such as a rude gesture. However, for the driver who was honked at to feel punished, it is not always necessary for the driver who honked at them to have intended to sanction them. The driver who was honked at, even if it was just to alert them, may still feel punished by the interaction. The critical point is that the overall pattern of honking exerts its influence on collective driving behavior via its inducement of individuals to change how they drive.

As you drive along, any time another driver is in hearing range of your horn constitutes a \emph{sanctioning opportunity}; you have an opportunity to honk your horn at a nearby driver and either chooses to do so or not to do so. Each time you honk the horn this constitutes a \emph{disapproval event} and each time-step when you do not honk is an \emph{approval event}. The \emph{context} of the sanctioning opportunity could be the current time at the point of sanctioning, or it could also optionally include some number of time-steps that preceded the sanctioning opportunity. While the sanction opportunities only occur if agents are within hearing distance, the sanction-observation function $\mathcal{B}$ can be either local or global. In the local case, an agent is only aware of a sanction opportunity and its outcome if it physically observed / experienced it. In the global case, we can imagine that $\mathcal{B}$ is streamed to a database and available to all agents. As an instantiation, one could image a dash-cam and microphone streaming every sanctioning opportunity and approval / disapproval to a database that would be accessible to all drivers and agents. This latter variant, in which all sanctioning opportunities and outcomes in an episode are available to all agents, is the main setting we consider in this work.

\subsection{Learning to classify transgression}

In this work we are concerned with developing a multi-agent simulation model where social norms emerge as the system self-organizes by learning. As such, the things the agents do in their world do not have any objective normative status. The classification of whether or not a given behavior constitutes a transgression is determined entirely by whether the group has sanctioned similar behavior in the past.

Each \emph{Classifier Norm Model} (\textbf{CNM}) agent has its own representation for what it thinks the group would sanction---i.e., a classifier that predicts whether the group would approve or disapprove of any given behavior. We train each individual's classifier on the public sanctioning observations provided by $\mathcal{B}(s_{0:T})$. Given a classifier $\Psi_\phi$ that outputs probabilities of sanctioning and assuming the set of sanctioning opportunities is of size $M$, we form a binary cross-entropy loss
\begin{equation*}
    \mathcal{L}_\phi = \frac{1}{M} \sum_{c, z \in \mathcal{B}} -z \log \left(\Psi_\phi(c)\right)  - (1 - z) \log\left(1 - \Psi_\phi(c) \right) 
\end{equation*}
and minimize it with stochastic gradient descent. 

There are some potential challenges with learning this classifier. One key issue arises because the classification is learned from the stream produced by an ongoing simulation. The data distribution may not be stationary. For example, when a particular behavior becomes effectively suppressed, perhaps because it was being punished so all agents learned to stop doing it, then the classifier will no longer receive training samples of it being approved or disapproved. This shift in the data distribution violates a stationarity assumption underpinning the classifier's training procedure and as a result, may cause \emph{catastrophic forgetting}~\citep{mcclelland1995there}, a phenomenon where a neural network unlearns its prior pattern of behavior.
To avoid this problem, we stop the classifier from continuing to learn after some fixed number of time-steps by setting its learning rate to zero. This freezes at that point in time each agents' representation of how context determines whether one has or has not transgressed, but it does not prevent subsequent drift in their sanctioning behavior or compliance behavior. 

\subsection{Learning how to enforce and comply}

The core idea of the \textbf{CNM} agent is that an individual embedded in a wider group is motivated to sanction in accord with the group's joint pattern of approval and disapproval. This shapes the group's behavior because disapproval is punishing.

The motivation to sanction consistently with the group is created by a pseudoreward term in the agent's reward function (i.e., an intrinsic motivation in the sense of \cite{singh2004intrinsically}) that encourages each reinforcement learning agent to disapprove in contexts that their classifier assesses as likely to provoke disapproval from others in the group:
\begin{equation*}
    \Omega_\phi(o_t, a_t) =
 \begin{cases} 
    +\alpha &\mbox{if } a_t \textrm{ is disapproval} \land \Psi_\phi(o_t) \ge 0.5 \\
    -\beta &\mbox{if } a_t \textrm{ is disapproval} \land \Psi_\phi(o_t) < 0.5 \\
    0 &\mbox{otherwise}
\end{cases}
\end{equation*}
for $\alpha, \beta \in \R^+_0$.

A \textbf{CNM} agent learns its classifier while simultaneously learning to maximize reward augmented by this intrinsic motivation to align its sanctioning with that of its group. It learns by applying a decentralized multi-agent reinforcement learning algorithm. Achieving high intrinsic reward demands the agent learn an efficient enforcement policy that sanctions like the wider group. Achieving high extrinsic reward demands the agent learn an efficient compliance policy that avoids provoking disapproval from others.

Each agent $i$ learns a parameterized behavior policy that is conditioned solely only the history of its own individual observations and actions and its estimate of the collective sanctioning pattern
$\pi_{\theta}(a^{(i)}_t|o^{(i)}_{0:t}, a^{(i)}_{0:t-1}, p_t)$ 
where $p_t = \text{stop}\left(\ind \left[\Psi_\phi(o_t) \ge 0.5\right] \right)$ 
and $\text{\emph{stop}}(\cdot)$ is the stop gradient operator.

\begin{figure*}
    \centering
    \includegraphics[width=0.8\textwidth]{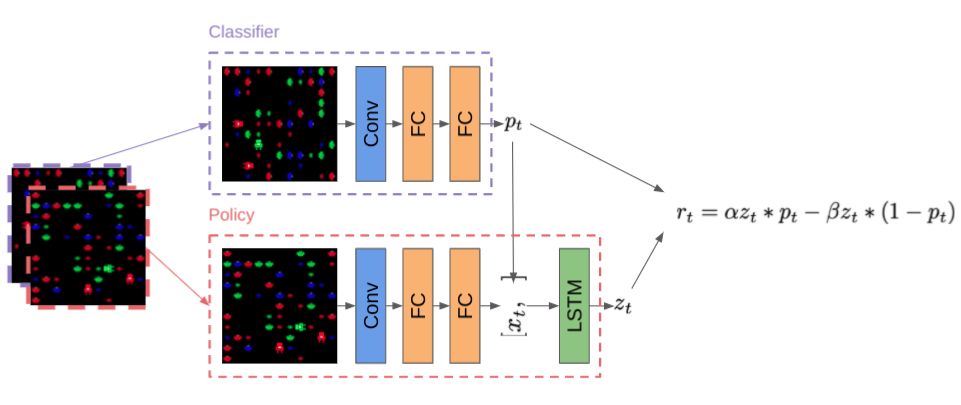}
    \caption{Visual depiction of how the classification is done and how the pseudoreward for aligning with the classifier is generated. The frame at which disapproval occurs and the frame before are stacked together; the frame before the disapproval is fed into the classifier to generate a prediction. If the agent chooses to disapprove, then a reward or penalty is generated based on whether its choice aligns with its classifier prediction.}
    \label{fig:class_architecture_2}
\end{figure*}

Both classifier and policy consist of a convolutional backbone attached to a multi-layer perceptron (MLP). The classifier MLP directly outputs the predictions whereas the policy MLP feeds into a recurrent network (an LSTM~\citep{hochreiter1997long}) whose outputs are the action probabilities. The classifier network takes the prior frame to make its prediction (context length is one, see Sec.~\ref{sec:sanction_obs}) whereas the policy takes the current frame to get an action. 
The classifier and policy do not share any layers in this architecture. The overall architecture, including the manner in which predictions are passed to the policy and the pseudoreward computation, is illustrated in Fig.~\ref{fig:class_architecture_2}. 

Each agent's policy is implemented using a private neural network, with no parameter sharing between agents. Each agent's policy parameters are independently trained to maximize the policy's long-term $\gamma$-discounted payoff:
\begin{align*} 
\label{eqn:policyobj}
V_{\theta, \phi}^{\vec{\pi}}(s_0) = \mathbb{E}_{\vec{\pi}_t, \mathcal{T}} \bigg[&\sum \limits_{t=0}^{\infty} \gamma^t \mathcal{R}_i(s_t, \vec{a}_t, s_{t+1}) 
+ \gamma^t \Omega_\phi(o^{(i)}_{t-1}, a^{(i)}_{t-1})\bigg]
\end{align*}

where the pseudoreward term shapes sanctioning behavior towards coherence with the group's pattern of approval and disapproval. We train on episodes sampled from $\vec{\pi}$. All agents control exactly one player in every episode.

The reinforcement-learning algorithm used for each agent is A3C~\citep{mnih2016asynchronous} with a V-Trace loss for computing the advantage~\citep{espeholt2018impala}. To the standard A3C loss we add a contrastive predictive coding loss \citep{oord2018representation} in the manner of an auxiliary objective \citep{jaderberg2017reinforcement}, which promotes discrimination between nearby timepoints via LSTM state representations. For more details please refer to the Appendix.

\section{Environments}\label{section:environments}

\begin{figure}
    \centering
    \includegraphics[width=0.4\textwidth]{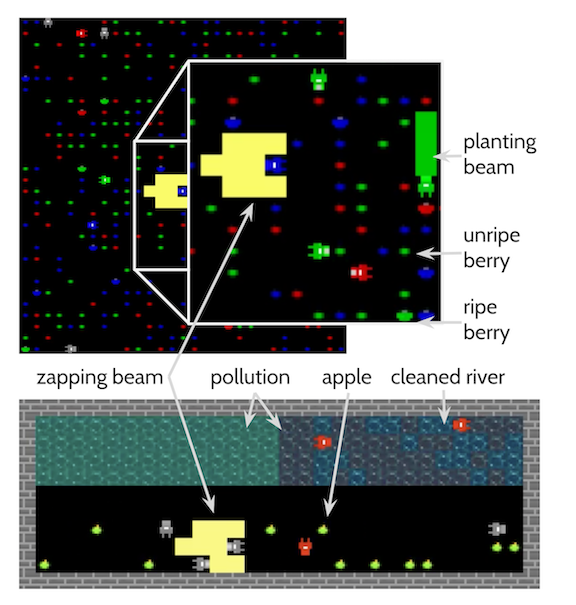}
    \caption{
    (\textbf{Top}) Allelopathic Harvest. Agents can recolor (replant) berries using one of three colored beams; a green beam is shown here. An agent's color is given by the berry color they most recently changed a berry to be (planted) or stochastically reverts to gray upon eating a berry. They also can zap agents to punish them (yellow beam).
    \\
    (\textbf{Bottom}) Clean Up with Startup Problem. Agents have a cleaning beam that can be used to clean pollution on either side of the divide as well as having a zapping beam that they can use to punish agents.}
    \label{fig:games_figure}
\end{figure}

We study two complex collective action problems implemented in Melting Pot~\citep{leibo2021scalable}. The two games are depicted in Fig.~\ref{fig:games_figure}, \emph{Allelopathic Harvest}\footnote{See  \textcolor{blue}{\url{https://youtu.be/la24sFmk6l8}} and \textcolor{blue}{\url{https://youtu.be/A4zMh9359r8}} for videos of example episodes of \emph{AH} and \emph{CSP}, respectively.} (\emph{AH}) and \emph{Clean Up with Startup Problem} (\emph{CSP}). Both games have the flavor of bargaining problems in the sense that several different Pareto-optimal outcomes are possible but individuals' preferences over said outcomes conflict with one another. Both games contain several different equilibria, each associated with a distinct type of ``work'' and superior to other uncoordinated equilibria. Thus both games contain start-up and free-rider sub-problems (terminology from~\cite{marwell1993critical}). This means that in order to achieve high rewards the agents must distribute some amount of work among themselves (cooperate) and most of that work should advance the same unified goal (coordinate). Learning in both games may be decomposed loosely into two phases. First, before much learning has occurred, very few individuals work consistently toward any goal so defection is motivated by fear that too few others will contribute to successfully establish any norm (the start-up problem). In the later phase of learning, when most individuals are engaged, then the motivation to defect is greed since one can free-ride on the efforts of others~\citep{heckathorn1996dynamics}. Games with this kind of bargaining-like collective action problem structure were previously studied with MARL in~\cite{koster2020model}.

In \emph{Allelopathic Harvest} (adapted from~\cite{koster2020model}), agents are presented with an environment that contains three different varieties of berry (red, green, and blue) and a fixed number of berry patches, which can be replanted to grow any color variety of berry. The growth rate of each berry variety depends linearly on the fraction that that variety (color) comprises of the total. As depicted in Fig.~\ref{fig:games_figure}, agents have three planting actions with which they can replant berries in front of themselves in their chosen color. Agents in AH have heterogeneous tastes. Specifically, half the agents receive twice as much reward from eating red berries relative to other berries and the other half have preferences of the same form except that they favor green. Agents can achieve higher return by selecting just one single color of berry to plant, but which one to pick is difficult to coordinate (start-up problem). They also always prefer to eat berries over spending time planting (free-rider problem).

In \emph{Clean Up with Startup Problem} (adapted from~\cite{hughes2018inequity}), the agents need to coordinate on a specific type of pollution to clean out of two pollution types as is shown in Fig.~\ref{fig:games_figure}. The environment contains apples that the agents are rewarded for eating, but the apple spawn rate increases monotonically with the ratio between the two pollution types. If the agents clean both pollution types equally, then apples will not spawn at all. Agents thus need to coordinate on a particular pollution type to clean (start-up problem) while also incentivizing enough agents to do the work of cleaning (free-rider problem).

Both environments have a rule with an effect similar to the cookie example from Sec.~\ref{sec:sanction_obs}. Individuals can see which kind of work (or free riding) other individuals have recently been engaged in. They change color to reflect this information. This makes it easier for agents to identify free-riders and those planting prohibited berry varieties (AH) or cleaning the wrong kind of pollution (CSP). In both environments the agents are colored according to their most recent planting or cleaning action. For example, successful planting of a red berry (AH) or successful cleaning of red pollution (CSP) causes the agent itself to become red. Similarly, agents that eat fruit are colored grey to indicate that they have not recently planted or cleaned. Thus grey colored agents are typically free riding.

In both environments agents can zap one another at short-range with a beam. This serves as the punishment mechanism. Importantly, in both games there are also instrumental reasons for agents to zap one another, especially to compete for  berries/apples. Getting zapped once freezes the zapped agent for 25 steps and applies a mark that indicates that the agent did something that was disapproved of (similar to~\cite{koster2022spurious}). If a second zap is received while the agent is marked, the agent is removed for 25 steps and receives a penalty of $-10$. If no zap is received for 50 steps, the mark fades. For full details on the environment please refer to Appendix Sec. B.

\section{Experiments}

\subsection{Existence and Beneficial Effects of the Emergent Social Norms:}

\begin{figure}
    \centering
    \includegraphics[width=.4\textwidth]{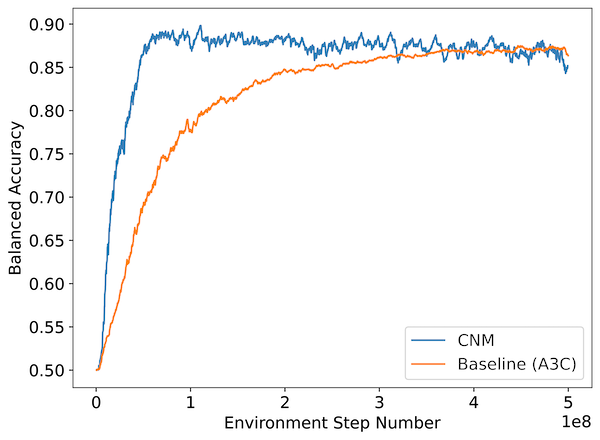}
    \caption{The classifier achieves high balanced accuracy (the average accuracy over both positive and negative samples) in predicting approval versus disapproval events.}
    \label{fig:balanced_acc}
    
\end{figure}

In order to align themselves with the social norm, agents must first learn to represent it accurately. Fig.~\ref{fig:balanced_acc} shows the balanced accuracy of the classifier in two cases where pseudorewards are on and one where the classifier is left on but has no influence in the environment. We observe three features. First, we are able to rapidly learn a classifier that achieves high balanced accuracy. Our ability to achieve high accuracy despite using only a single frame suggests that the initial normative behavior is something simple like "zap an agent if it might compete with you over a visible berry" or "zap agents of a particular color. Second, we note that the pseudorewards from the classifier in turn cause the accuracy of the classifier to rapidly converge; the agents adjust their behavior to be in accord with the classifier. Finally, we freeze the classifier after $5e7$ steps, but despite this the balanced accuracy remains relatively high for the duration of training, suggesting that there is not too much drift in the norm after the freeze. Similar behavior is observed in \emph{CSP}.

\begin{figure*}
\centering
\begin{subfigure}{.24\textwidth}
  \centering
  \includegraphics[width=\linewidth]{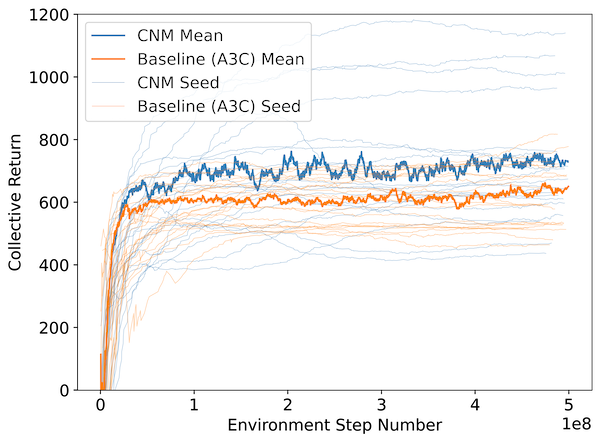}  
  \caption{}
  \label{fig:collective_allelo}
\end{subfigure}
\begin{subfigure}{.24\textwidth}
  \centering
\includegraphics[width=\linewidth]{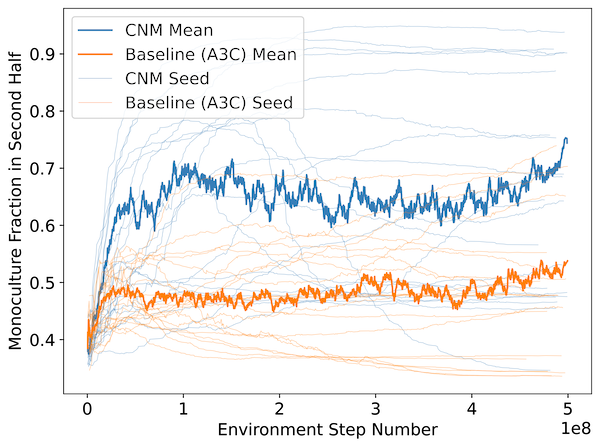}  
  \caption{}
  \label{fig:monoculture_allelo}
\end{subfigure}
\begin{subfigure}{.24\textwidth}
  \centering
  \includegraphics[width=\linewidth]{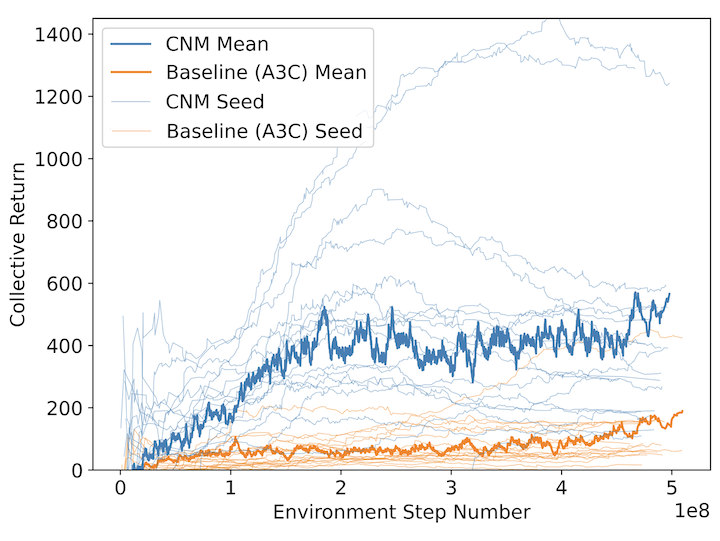}  
  \caption{}
  \label{fig:collective_cleanup}
\end{subfigure}
\begin{subfigure}{.24\textwidth}
  \centering
  \includegraphics[width=\linewidth]{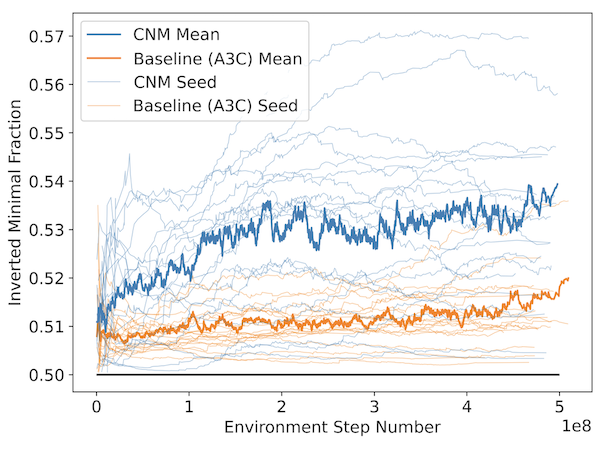}  
  \caption{}
  \label{fig:inverted_cleanup}
\end{subfigure}
        \caption{The effect of norms on avoiding startup problems and overcoming freerider problems. The thick lines represent the mean across seeds while thin, transparent lines represent individual seeds; the standard deviation is not displayed for visual clarity.
        (a) Collective return in \emph{AH}. (b) Fraction of total berries constituted by the dominant berry in the second half of the episode. (c) Collective return in \emph{CSP}. (d) Average fraction of total pollution constituted by the dominant pollution type.}
        \label{fig:allelo_monoculture}
\end{figure*}

Next, we investigate whether the use of \textbf{CNM} leads to better outcomes. In \emph{AH} we run 20 seeds and in \emph{CSP} we run 10 seeds. In \emph{AH}, the measure of success is the \emph{monoculture fraction}, the percentage of the color that corresponds to the largest number of berry spawning sites. Fig.~\ref{fig:monoculture_allelo} demonstrates that \textbf{CNM} increases the monoculture fraction above 50\%, indicating that agents on average are converging to a single preferred color, and also increases the net agent return, indicating that the costs of norm enforcement (punishing violators) are overcome by increased berry consumption. Similarly, we observe that in \emph{CSP} they are able to successfully select one of the two pollution types over the other. The inverted minimal fraction measures how imbalanced the two types of pollution are; higher inverted minimal fraction is desirable. The result is a significant consequent increase in collective return. Note that collective return, as defined here, includes the costs of being punished since these are externally imposed by other agents but does not include the pseudoreward term since it models an internal drive.

Groups of \textbf{CNM} agents display a bandwagon effect, magnifying weak patterns of sanctioning in initially random exploratory behavior. They are more likely than the baseline to coordinate on a coherent joint behavior (planting a specific berry color in AH or cleaning a specific pollution type in CSP). But there is no guarantee that they will select the most beneficial equilibria available to them. This mirrors the arbitrariness of real-world social norms. For example, recall that all agents in AH prefer either red or green berries over blue berries (see Sec.~\ref{section:environments}). If agents have an early tendency to plant the undesirable blue berries and punish free-riders, the classifier will learn to approve of these behaviors and the agents will stabilize on a  blue equilibrium, an outcome that none of them prefer over red or green equilibria. This is why there is so much variation in the outcomes achieved between independent runs (Fig.~\ref{fig:allelo_monoculture}). See also Fig.~\ref{fig:allelo_simplex} where the prevalence of blue berry centric outcomes can clearly be seen.

\begin{figure}
\centering
\begin{subfigure}{.4\textwidth}
  \centering
  \includegraphics[width=\linewidth]{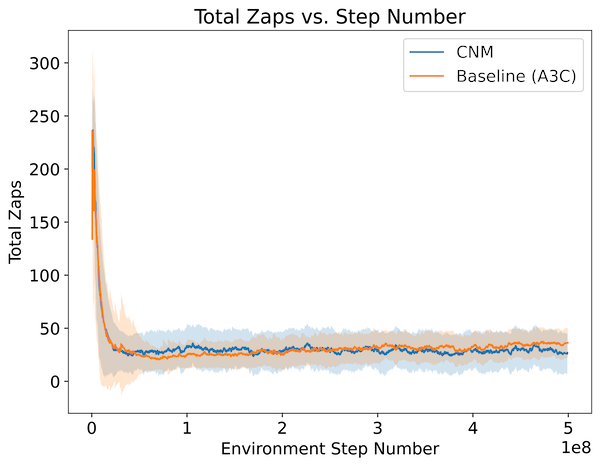}  
  \caption{}
  \label{fig:ah_zaps}
\end{subfigure}
\begin{subfigure}{.4\textwidth}
  \centering
  \includegraphics[width=\linewidth]{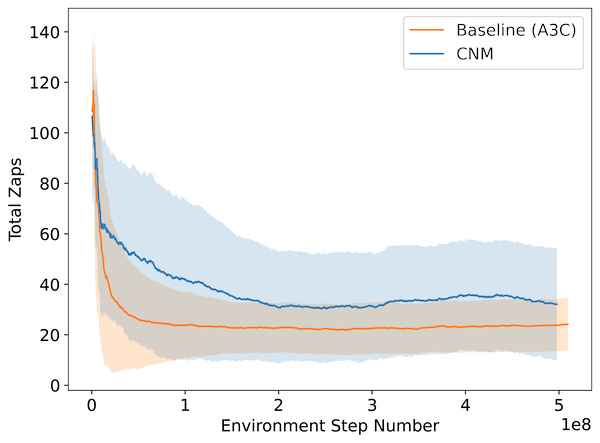}  
  \caption{}
  \label{fig:csp_zaps}
\end{subfigure}
\caption{Effect of CNM on total number of zaps averaged across seeds in (a) \emph{AH} (b) \emph{CSP}.}
\label{fig:total_zaps}
\end{figure}

Finally, we confirm that the improvement in reward is not somehow occurring due to a suppression of the penalty action and a consequent decrease in penalty from zap events; rather, the total amount of punishment events actually stays the same or even increases with \textbf{CNM}. Remember, zapping can also be used instrumentally, e.g., to compete over berries or apples. Fig.~\ref{fig:total_zaps} shows the average number of zaps in an episode summed over the agents for \emph{AH} and \emph{CSP}. Note that there is no observable amount of difference in the net amount of zapping for \emph{AH} and zapping increases for \emph{CSP}. Thus, improvements in collective return must be coming from changes in how zapping is used.

\subsection{How does \textbf{CNM} establish social norms?}

\begin{figure*}
\centering
\begin{subfigure}{.24\textwidth}
  \centering
  \includegraphics[width=\linewidth]{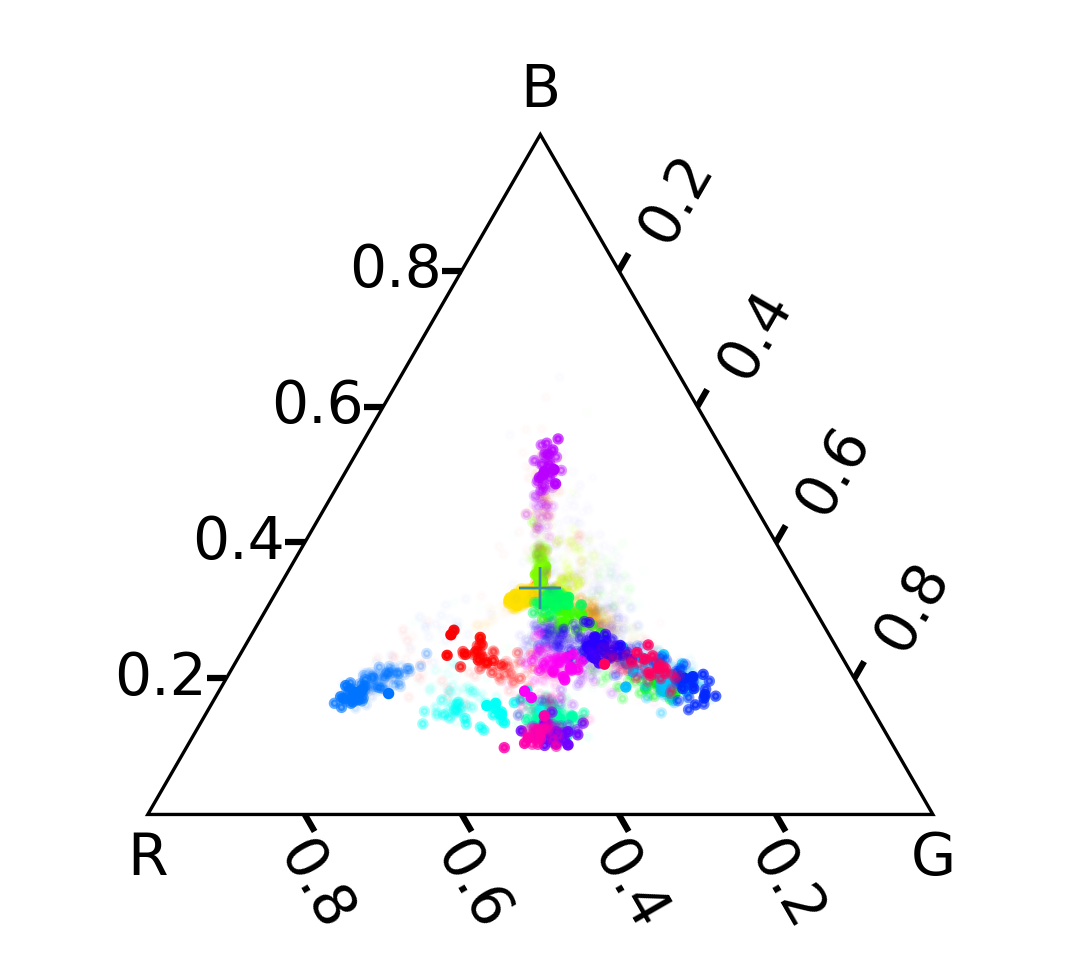}  
  \caption{}
\end{subfigure}
\begin{subfigure}{.24\textwidth}
  \centering
  \includegraphics[width=\linewidth]{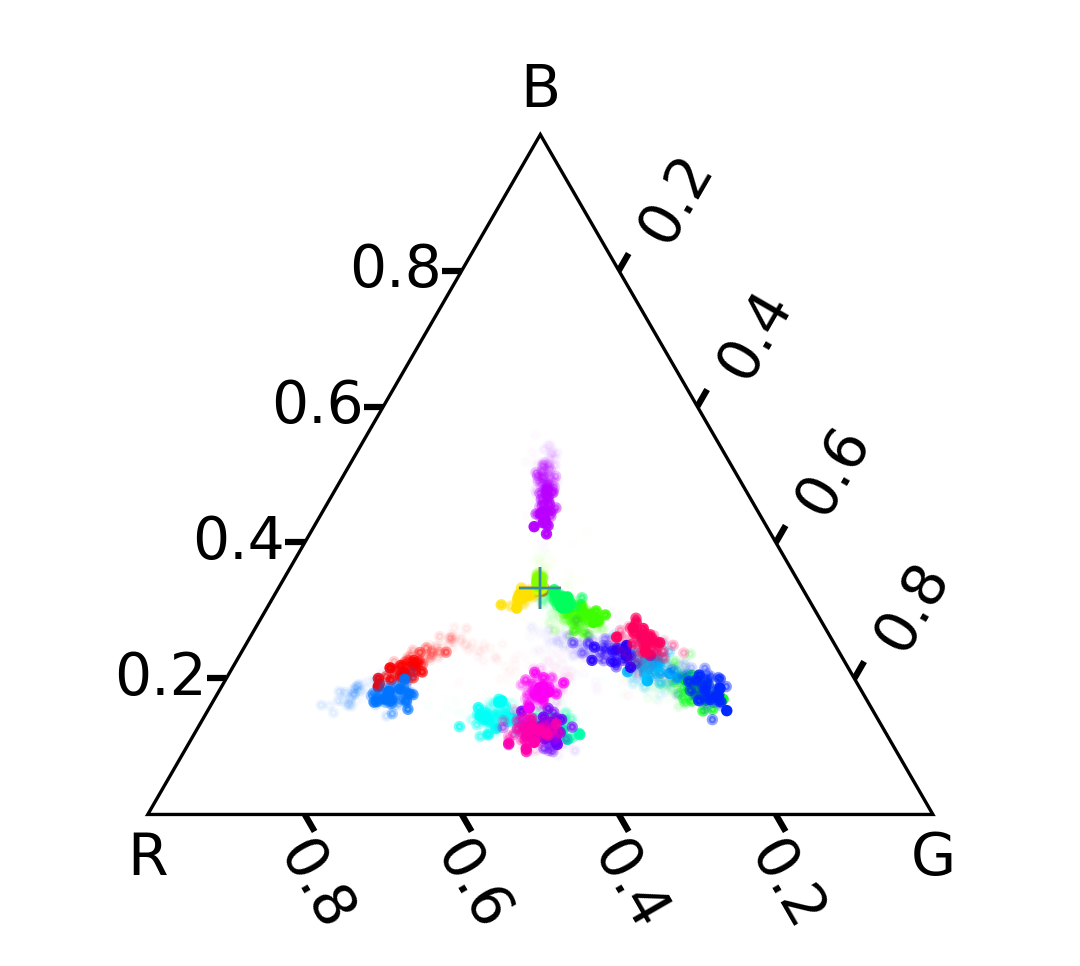}  
  \caption{}
\end{subfigure}
\begin{subfigure}{.24\textwidth}
  \centering
  \includegraphics[width=\linewidth]{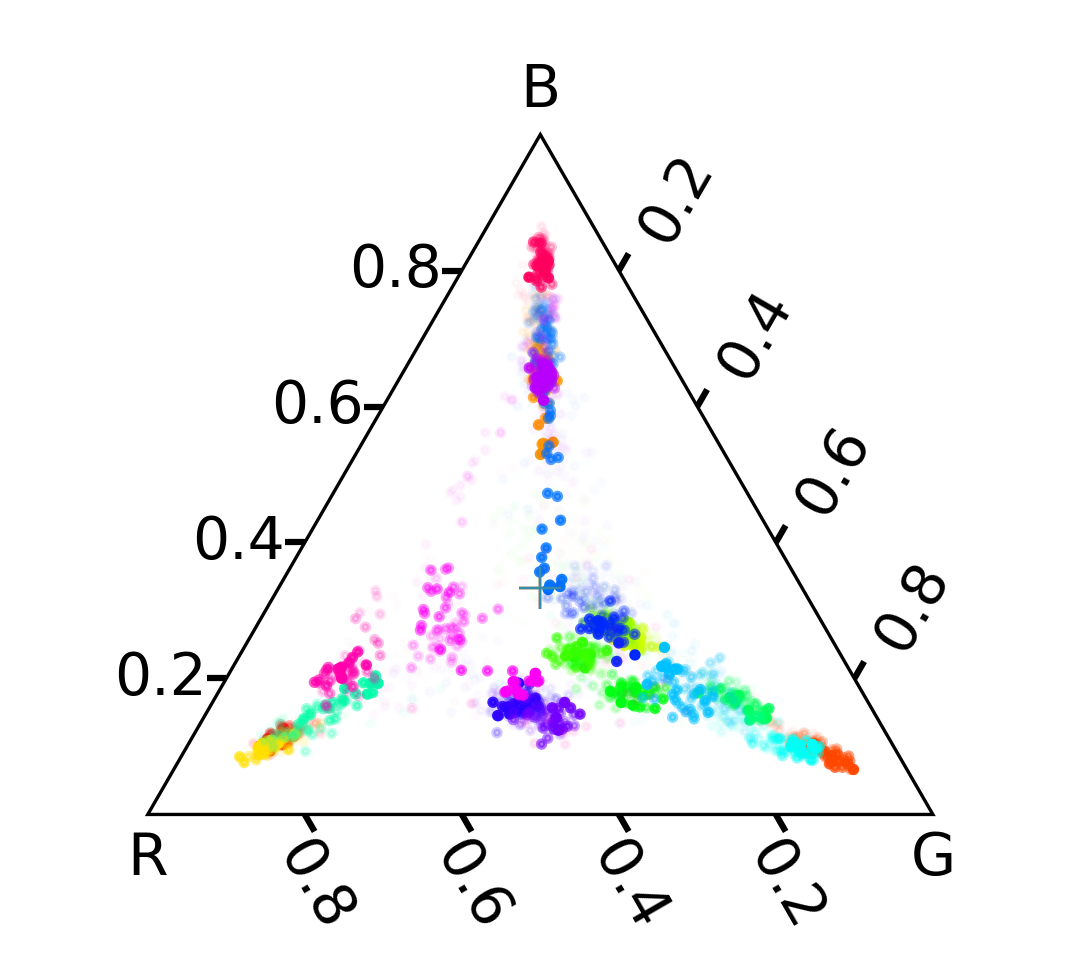}  
  \caption{}
\end{subfigure}
\begin{subfigure}{.24\textwidth}
  \centering
  \includegraphics[width=\linewidth]{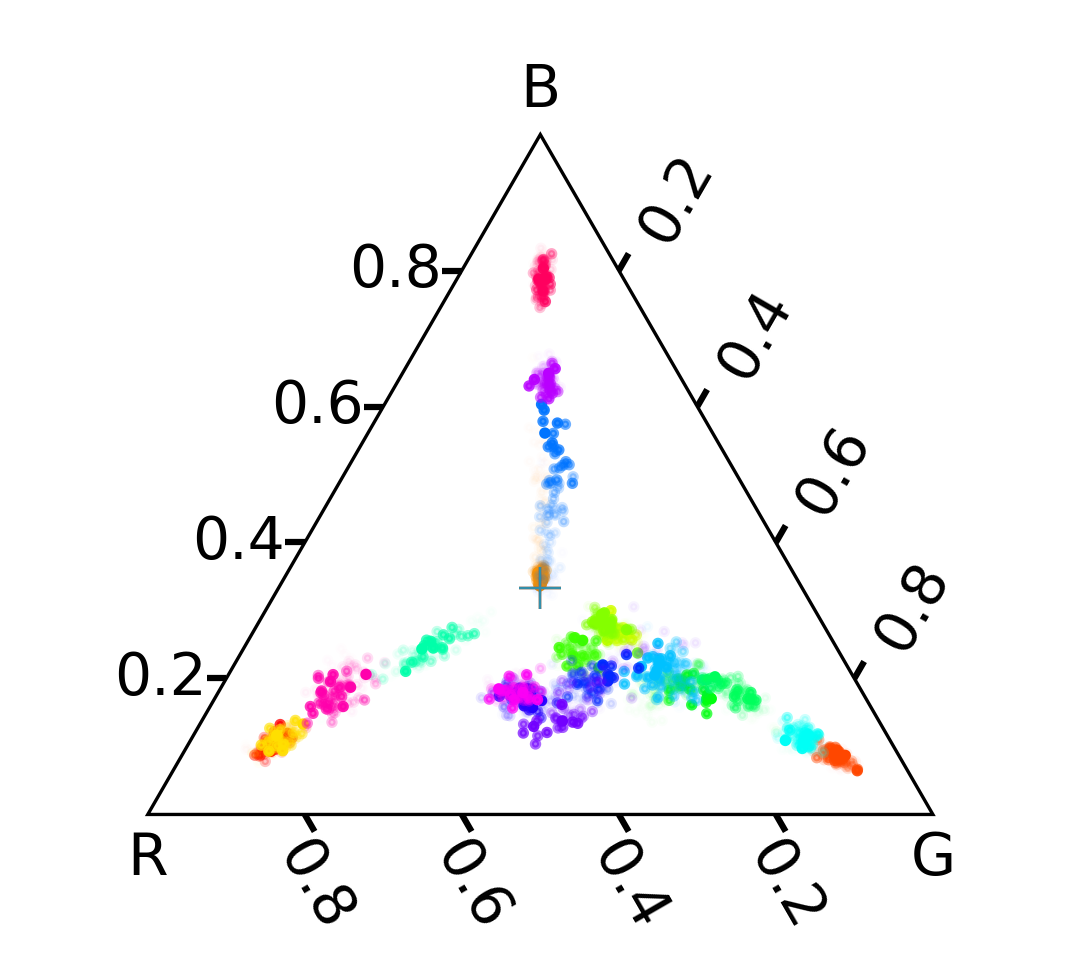}  
  \caption{}
\end{subfigure}

    \caption{Evidence of stable planting behavior after $2e8$ steps of training. Individual dots are samples over a run where darker dots represent later points. (a) First $2e8$ steps with \textbf{CNM} off. (b) Latter $2.5e8$ steps with \textbf{CNM} off. (c) First $2e8$ steps with \textbf{CNM} on. (d) Latter $2.5e8$ steps with \textbf{CNM} on.}
    \label{fig:allelo_simplex}

\end{figure*}

Here we show that \textbf{CNM} increases incentives to obey social norms i.e. agents are disapproved of more for deviating from the established equilibrium. In \emph{AH}, the equilibria are likely given by the corners of the berry fraction simplex (Fig.~\ref{fig:allelo_simplex}). Stabilization comes from disapproval of re-planting behaviors that would push away from an equilibrium. We can approximately observe stability in the planting behavior by examining the evolution of the fraction of each berry color on the simplex. Fig.~\ref{fig:allelo_simplex} demonstrates the changes in the evolution of berry fraction during early and late phases of training. Here the center of the diagram indicates that either all agents are free-riding or that they are all cancelling out one another's planting behavior (e.g.~I change a red berry to blue and you change a blue berry to red so there is no net effect on the berry fractions). We observe that groups of \textbf{CNM} agents push further away from the center and towards the corners of the simplex. Furthermore, there is little change in later steps of training for the seeds that reach the simplex corners, suggesting an equilibrium. There is some small amount of drift in high blue monoculture fractions which may be occurring as the blue berries are not preferred by any agent.
\begin{figure}
\centering
\begin{subfigure}{.4\textwidth}
  \centering
  \includegraphics[width=\linewidth]{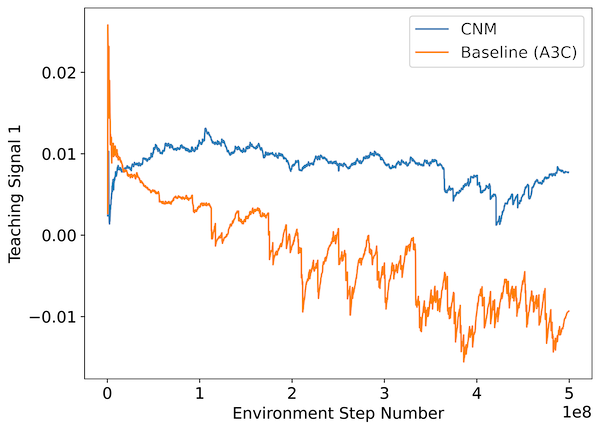}  
  \caption{}
  \label{fig:allelo_gray_zap}
\end{subfigure}
\begin{subfigure}{.4\textwidth}
  \centering
  \includegraphics[width=\linewidth]{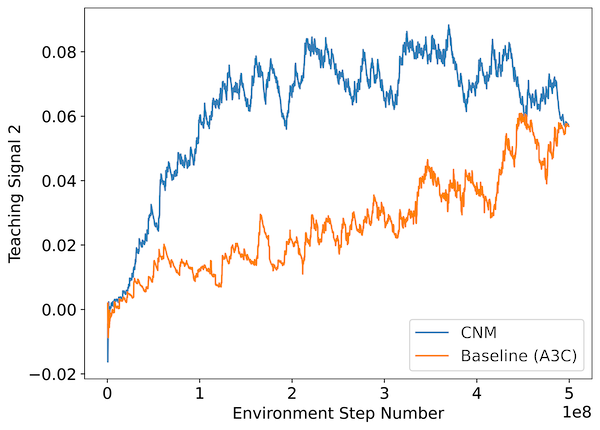}  
  \caption{}
  \label{fig:allelo_dominant_zaps}
\end{subfigure}
\caption{Measurements of the strength of disapproving sanctions applied to deviating agents in allelopathic harvest: (a) difference between zap likelihood for being the free-rider color vs. being the dominant color (b) difference of zap likelihood for being the second most dominant color to being the dominant color.}
\label{fig:sanctioning_behavior}
\end{figure}

The second criterion to check concerning the establishment of a social norm is that deviations from the equilibrium should be disapproved (sanctioned). We can calculate for each color $\text{\textbf{p(zapped | color)}}$ by Bayes' rule (details in Appendix). We then use it to investigate the sanctioning forces supporting a particular equilibrium by looking at the difference in log likelihood of being punished while working toward establishing or maintaining the equilibrium. Agents can readily perceive which equilibrium other agents in their field of view are supporting because their color shows which color berry they last planted (see Section~\ref{section:environments}). If the likelihood difference for a particular color is high it should be easy for the learning algorithm to identify that switching to that color (i.e.~switching to support its corresponding equilibrium) is likely to lead to disapproval. Thus, these differences serve as a teaching signal pushing the agent towards planting one color and away from planting another.

Fig.~\ref{fig:sanctioning_behavior} demonstrates this effect for two different potential switches. Fig.~\ref{fig:allelo_gray_zap} measures the difference of punishment likelihood between free-riding and planting the dominant color which we call \emph{teaching signal 1}. If the magnitude of this signal is large and positive, it is easy for the learning algorithm to identify that switching from free-riding to planting in that color will decrease the amount that it gets punished.

Fig.~\ref{fig:allelo_dominant_zaps} measures the relative likelihood of getting punished when we plant the color corresponding to high monoculture versus if we were to switch to plant the second most abundant berry color which we refer to as \emph{teaching signal 2}. If this signal is large, it is easier for the learning algorithm to identify that sticking to the dominant color will allow it to decrease how often it gets disapproved of which in turn will help stabilize the choice of equilibrium. 

\subsection{Ablations on architecture components}
\label{sec:ablations}

To understand \textbf{CNM} better we gradually remove and alter components of the architecture to answer the following questions: (1) Is freezing the classifier necessary? (2) Is it essential to learn social norms from global sanctions or will local sanctions observed by each individual themselves suffice? (3) Is our result sensitive to the relative scale between approval and disapproval pseudorewards?

Here we study \emph{CSP} as the smaller number of agents in this environment decreases environment step time and allows us to perform more rapid experimentation. We run each ablation over ten seeds. For point (1), we allow the classifier to continue learning throughout training. For (2) we train the classifier using only the sanctioning events directly observed by each agent. Finally, for (3), we note that in all prior experiments we have scaled the pseudorewards so that the penalty for punishing discordantly with the classifier $(\beta)$ is twice the reward for punishing in accord with it $(\alpha)$. We aim to establish whether our results are sensitive to this particular ratio. 

\begin{figure}
\centering
\begin{subfigure}{.4\textwidth}
  \centering
  \includegraphics[width=\linewidth]{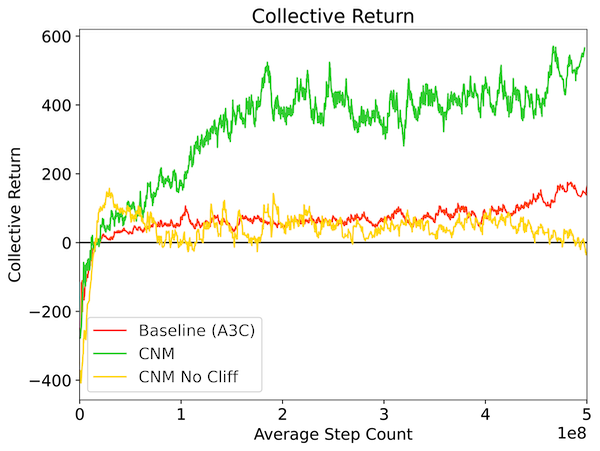}  
  \caption{}
  \label{fig:ablation_a}
\end{subfigure}
\begin{subfigure}{.4\textwidth}
  \centering
  \includegraphics[width=\linewidth]{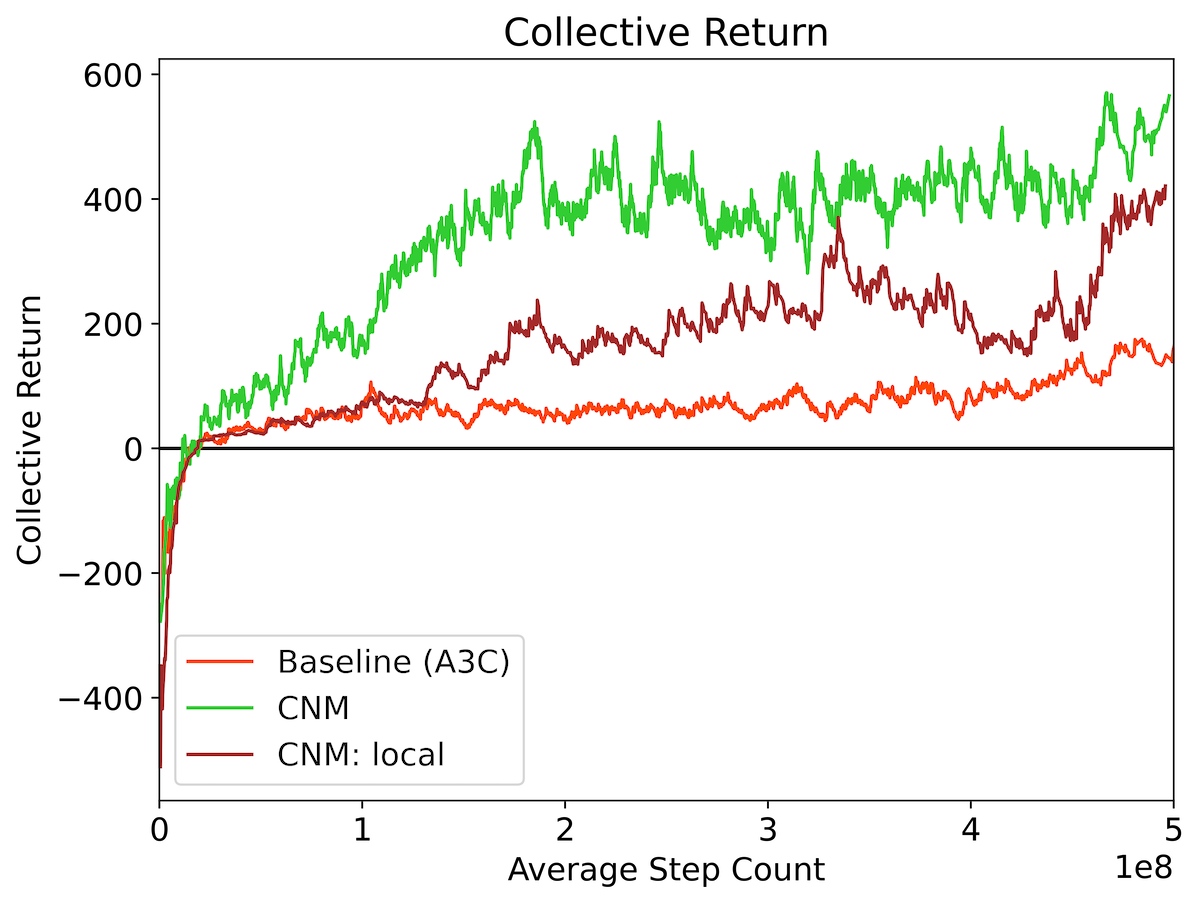}  
  \caption{}
  \label{fig:ablation_b}
\end{subfigure}
\begin{subfigure}{.4\textwidth}
  \centering
  \includegraphics[width=\linewidth]{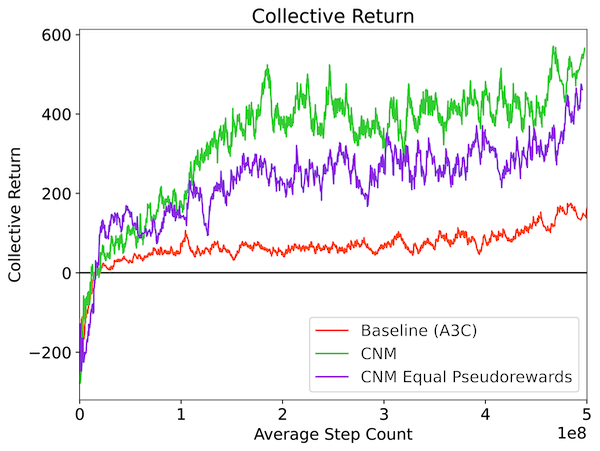}  
  \caption{}
  \label{fig:ablation_c}
\end{subfigure}
\caption{Ablations of key components of the agent architecture. (a) The classifier is not frozen during training. (b) The classifier is learned solely from sanctions experienced by the agent i.e sanctions are private. (c) Effect of pseudoreward scale; both $\alpha$ and $\beta$ are set to 0.9.}
\label{fig:ablations}
\end{figure}

Fig.~\ref{fig:ablations} demonstrates the outcome of all of these ablations; each curve is the average across ten seeds with std. deviations removed for visual clarity. In Fig.~\ref{fig:ablation_a} we can see that in the absence of a frozen classifier the collective return experiences a large early spike but then decays quickly down. While we are unable to definitively establish the mechanism that forces us to freeze the classifier, there are a few plausible ones. The move away from free-riding occurs rapidly in the first $1e8$ steps of training (see Appendix Sec. C). If the punishment behavior is not correspondingly suppressed quickly enough, agents performing cooperative behavior will still get punished due to exploratory noise and the classifier will consequently learn to recommend punishment of cooperative agents. Alternately, the classifier could simply experience \emph{catastrophic forgetting} once a particular color is effectively suppressed: it's difficult to remember how to sanction a behavior that no longer occurs. Consequently, the suppressed behavior is able to re-emerge.

In Fig.~\ref{fig:ablation_b} we observe that learning solely from local sanctions does improve over the baseline but does not completely match the performance of fully public sanctions. Since the agents have to infer the norm solely through agents they happen to interact with, the number of samples available for each classifier update decreases sharply which may make the subsequent learned norm noisier and harder to learn.
Finally, in Fig.~\ref{fig:ablation_c} we set the pseudorewards to a magnitude of $0.9$ for both approval and disapproval. We note that this is less than the potential reward of consuming an apple, making it feasible for an agent to zap discordantly to the recommendation of the classifier if doing so nets them an additional apple. We see that there is a slight reduction in the collective return but there remains an improvement over the A3C baseline.

\section{Discussion and Future Work}

Motivated by emerging challenges in deploying multi-agent systems, we introduce and formalize a new training regime for decentralized multi-agent systems in which all sanctions are publicly observable. In contrast to centralized training methods, this approach can be trained fully online without needing access to a simulator. It also may make it easier to satisfy privacy constraints since essential proprietary data like rewards and policies do not need to be shared to achieve coordination.

We observe that in this setting decentralized agents struggle to achieve cooperative behavior in the collective action problems posed by two environments that broadly model challenges of free-riding and equilibrium selection. Inspired by social norms, which humans communities often use to overcome such dilemmas, we introduce an agent architecture \textbf{CNM} that learns to classify and enforce social norms from experience. We show that groups of \textbf{CNM} agents converge on beneficial equilibria and are better at resolving free-rider problems than agents implementing a baseline algorithm. 

However, many open questions remain. The architecture used for the classifier, a convolutional network, relies on there being an identifiable visual cue that correlates with the behavior to be made normative. Thus it is restricted in the types of norms it can identify. An extended \textbf{CNM} architecture operating on snippets of video preceding each sanctioning event may allow for different social norms to emerge.

Furthermore, while we observe the appearance of seemingly stable, beneficial norms, we do not provide a complete mechanistic explanation of how this architecture selects and stabilizes equilibria. It is possible that there exist games where this architecture would exclusively select harmful norms or deeply unfair norms. From the standpoint of using \textbf{CNM} for social science modeling, this is a feature not a bug. In the real world, for every beneficial norm enabling collective action, there are hosts of unsavory norms (but see also~\cite{hadfield2019legible, koster2022spurious}). Moreover, we must not take for granted that social norms are always a desirable outcome for a multi-agent system. For instance, social norms impose a deadweight loss due to the effort needed to maintain them. Paying this cost may not always be worthwhile in all applications. Nevertheless, we believe that \textbf{CNM}, or a successor system, could eventually be employed fruitfully in a wide range of applications from social science modeling to real-world multi-agent systems where interfacing with human social norms is especially critical.

\clearpage
\bibliographystyle{ACM-Reference-Format} 
\bibliography{biblio}

\newpage
\appendix
\section{Architecture and Algorithmic Details}
\subsection{Architecture}
In the implementation of our agent architecture and algorithm we aimed to stick with configurations proposed in recent work~\citep{koster2022spurious}. We made sure that they use the same size ConvNets and LSTMs. We didn’t perform any tuning of hyper-parameters and used the ones provided in the original publications studying the environments used here.

The agent’s network consists of a ConvNet with two
layers with 16, 32 output channels, kernel shapes 8, 4, and
strides 8, 1 respectively. It is followed by an MLP with two
layers with 64 neurons each. All activation functions are
ReLU and both the ConvNet and the MLP have activations at their final layer. It is followed by an LSTM with 128 units. Policy
and baseline (for the critic) are produced by linear layers
connected to the output of LSTM. 

Our classifier network uses the same architecture as the agent for its ConvNet. However, its MLP is three layers, (64, 64, 2) with the final layer not having an activation applied to its output. A softmax is applied to the output of this MLP to get the predicted probabilities of not-punish and punish respectively where the first index of the output corresponds to the probability of not punishing. As before, all activations in the ConvNet and MLP are ReLUs. 

\subsection{Classifier Training}
For training our classifier, we use batches of data returned by A3C. Each episode is chunked into segments of length 100. For each of these segments, we extract out all the events where
\begin{itemize}
    \item An agent is able to zap (there is a cooldown period after each zap is used during which time the zap action is unavailable).
    \item Another agent is within shooting range.
\end{itemize}
For each of these events, we then look at the action of the agent in the subsequent time-step to acquire a label: 0 for no zap, 1 for zap. Since there are sixteen agents and all sanction events are global, we have up to $1600$ possible punishment events in a batch. From these events, we randomly subsample $P=32$ of the punishment events and $P'=1024$ of the events where no punishment occurred. If $p_i$ is the classifier output on event $i$ we then form the cross-entropy loss
\begin{equation}
    \mathcal{L_\text{class}} = \lambda_{\text{class}} \left(\frac{1}{P}\sum_i^P \log(p_i) + \frac{1}{P'}\sum_i^{P'}\log(1 - p_i) \right)  
\end{equation}
where $\lambda_\text{class}$ is a scaling factor used to adjust the learning rate of the classifier relative to the learning rate of A3C. The classifier is trained via RMSProp with hyperparameters given in Sec.~\ref{sec:hyperparams}.

\subsection{Motivation to align punishment with group}
\label{sec:pseudoreward}
Given a classifier, we then use its predictions to add a pseudoreward to batches of data returned by A3C. As before, we select all potential sanctioning events. We feed the frame before the sanctioning event to the classifier and generate a prediction. The frame on which the sanctioning event occurs is fed into the policy and the classifier prediction concatenated onto the policy internal state after the MLP and before the LSTM. If the policy outputs a zap action out of its LSTM, we receive a positive reward if the classifier predicted zap as well and a penalty if the classifier predicted not to zap. This process is depicted visually in Fig.~\ref{fig:class_architecture_2}.

\subsection{Algorithm}
In addition to the standard A3C loss with the advantages computed using V-Trace~\citep{espeholt2018impala},
we used an auxiliary
loss~\citep{jaderberg2017reinforcement} for shaping the representation
using contrastive predictive coding~\citep{oord2018representation} (CPC). CPC
here discriminates between nearby time points via LSTM
state representations (a standard augmentation in recent
works with A3C).

\noindent
\textbf{Contrastive Predictive Coding Loss:}

At a high level, CPC works by taking the input and output of an RNN and trying to predict future RNN outputs from the RNN inputs. It does this over several time-shifts and performs the prediction in a latent space. 

Let $q^i \in \mathbb{R}^{L \times B \times N}$ denote the input to the LSTM layer and  $q^o \in \mathbb{R}^{L \times B \times N'}$ the output of the RNN where $L$ is the length of the time-slice provided to A3C, $B$ is the batch and $N, \, N'$ are the dimensions of RNN input and output respectively. For notational convenience, we will write $q_{k:j}$ to denote slices of the matrix $q$ along the time axis i.e. $k:j$ denotes the k'th to j'th element along the time axis. 
We apply a 1-d convolution $C$ with a kernel of size $1$ to both input $q^i$ and $q^o$ to project them to a latent dimension of size $l$, $C q^i \in \mathbb{R}^{\left(L * B\right) \times l}$ where we have folded a matrix reshape into the convolution. 

Finally, let 
$\mathcal{L}_{cse}$ denote the softmax cross-entropy loss and where the loss between a matrix and a matrix will be understood to mean applying the loss element-wise  and then computing the mean. The CPC update can then be written as 
\begin{equation}
    \mathcal{L_\text{CPC}} = \frac{1}{S}\sum_{i=1}^S\mathcal{L}_{\text{cse}}\left(C q^i_{s:L} \left(C q^o_{0:L-s}\right)^T, \boldsymbol{I}^{(L - s) * B \times (L - s) * B} \right)
\end{equation}
Here $S$ represents that we do prediction over all possible time shifts from $1$ to $S$ and $T$ is the transpose operation. 
\section{Environment Details}
Here we define the environment dynamics in as much detail as possible.
First, a few details that are shared between the two environments.
In both environments we will refer to a \emph{grid cell}, which we will define as an (8,8) square of pixels. In both environments the agents have a position and 4 possible rotations that are indexed from 0 to 3 where $0$ is North, $1$ is East, $2$ is South, and $3$ is West. Agents can take movement actions where are defined with respect to their current rotation i.e. we have 4 actions Up, Left, Right, Down and when the rotation is $0$, Up will move you North whereas when the rotation is $1$, Up will move you East.

\begin{figure}
    \centering
    \includegraphics[scale=0.25]{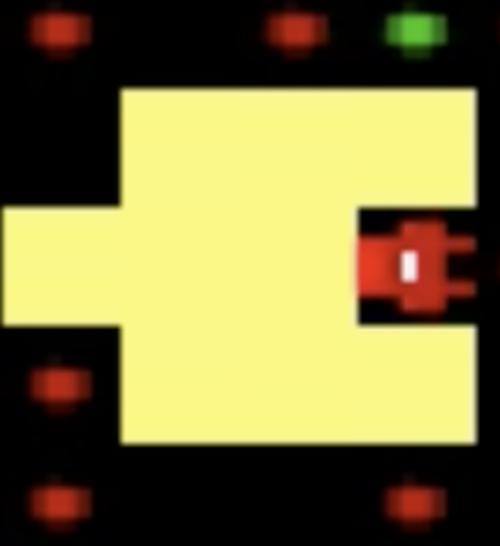}
    \caption{The zapping beam extended fully.}
    \label{fig:unblocked_zap}
\end{figure}

\begin{figure}
    \centering
    \includegraphics[scale=0.25]{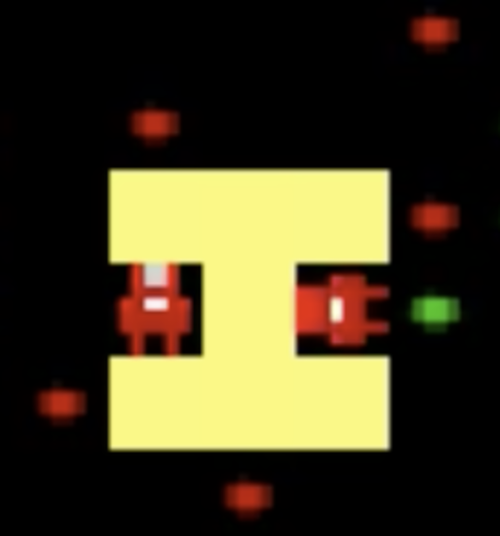}
    \caption{The zapping beam is blocked due to hitting an agent. }
    \label{fig:blocked_zap}
\end{figure}

Both environments contain a zapping beam that can be used to zap agents. Note that the use of this beam is an action that cannot be combined with other actions i.e. if an agent zaps then it cannot also move at that time-step. The zapping beam, as depicted in Fig. ~\ref{fig:unblocked_zap}, extends three grid cells forward from the direction the agent is facing. Additionally, on either side of the agent it also extends three grid cells forward but with the beam emanating from the grid cells directly next to the agent rather than from the grid cell directly in front of the agent. Agents hit by the beam block its forwards progress, as depicted in Fig.~\ref{fig:blocked_zap}.

Agents that are zapped by the grid cell acquire a marking, shown in Fig.~\ref{fig:blocked_zap}, and are frozen for 25 time-steps. In the frozen state they cannot perform any actions for that duration. If they are zapped again within 50 time-steps, the agent that is zapped receives a reward of $-10$ (i.e. a penalty) and is removed from the environment for 25 timesteps. During this time, all image based observations will be replaced with an image that is entirely black. 

\begin{figure}
    \centering
    \includegraphics[width=0.4\textwidth]{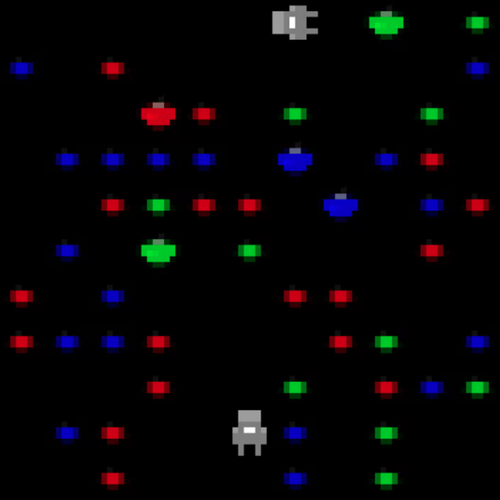}
        \includegraphics[width=0.4\textwidth]{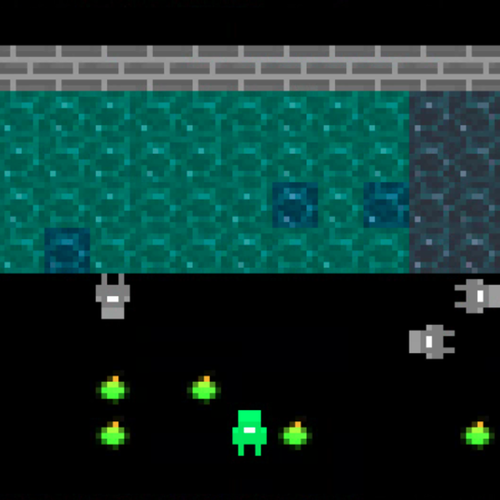}
    \caption{(Left) Observation of an agent in allelopathic harvest. (Right) Observation of an agent in cleanup.}
    \label{fig:local_view}
\end{figure}

In both environments the observation provided to the policy is an (88, 88, 3) RGB image that is centered on the agent as depicted in Fig.~\ref{fig:local_view} as well as the prediction of the classifier if the classifier is used as discussed in section~\ref{sec:pseudoreward}. The agent sees 9 grid cells in front of it, 1 grid cell behind it, and 5 grid cells to its left and right. Note that "in front, behind, left, right" are all defined with respect to the current rotation of the agent. Cells that fall outside the boundaries of the environment (since the world map is of finite size in Cleanup With Startup Problem) are returned as black.

\subsection{Allelopathic Harvest}
\subsection{Initial Map}
At environment reset, the environment is set to the following settings 

\begin{figure}
    \centering
    \includegraphics[width=0.4\textwidth]{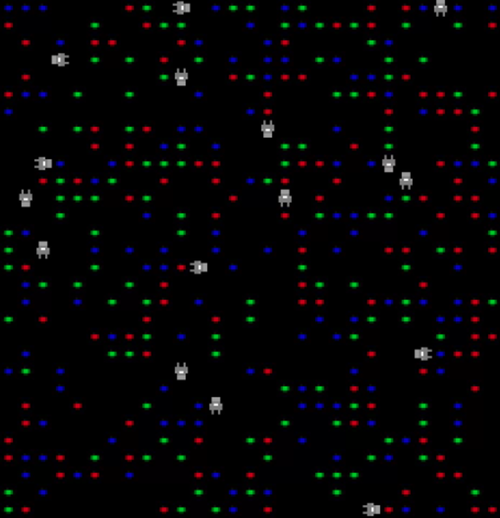}
    \caption{Allelopathic Harvest map at the first time-step.}
    \label{fig:allelo_reset}
\end{figure}

\noindent
\texttt{333PPPP12PPP322P32PPP1P13P3P3 \\
1PPPP2PP122PPP3P232121P2PP2P1 \\
P1P3P11PPP13PPP31PPPP23PPPPPP \\
PPPPP2P2P1P2P3P33P23PP2P2PPPP \\
P1PPPPPPP2PPP12311PP3321PPPPP \\
133P2PP2PPP3PPP1PPP2213P112P1 \\
3PPPPPPPPPPPPP31PPPPPP1P3112P \\
PP2P21P21P33PPPPPPP3PP2PPPP1P \\
PPPPP1P1P32P3PPP22PP1P2PPPP2P \\
PPP3PP3122211PPP2113P3PPP1332 \\
PP12132PP1PP1P321PP1PPPPPP1P3 \\
PPP222P12PPPP1PPPP1PPP321P11P \\
PPP2PPPP3P2P1PPP1P23322PP1P13 \\
23PPP2PPPP2P3PPPP3PP3PPP3PPP2 \\
2PPPP3P3P3PP3PP3P1P3PP11P21P1 \\
21PPP2PP331PP3PPP2PPPPP2PP3PP \\
P32P2PP2P1PPPPPPP12P2PPP1PPPP \\
P3PP3P2P21P3PP2PP11PP1323P312 \\
2P1PPPPP1PPP1P2PPP3P32P2P331P \\
PPPPP1312P3P2PPPP3P32PPPP2P11 \\
P3PPPP221PPP2PPPPPPPP1PPP311P \\
32P3PPPPPPPPPP31PPPP3PPP13PPP \\
PPP3PPPPP3PPPPPP232P13PPPPP1P \\
P1PP1PPP2PP3PPPPP33321PP2P3PP \\
P13PPPP1P333PPPP2PP213PP2P3PP \\
1PPPPP3PP2P1PP21P3PPPP231P2PP \\
1331P2P12P2PPPP2PPP3P23P21PPP \\
P3P131P3PPP13P1PPP222PPPP11PP \\
2P3PPPPPPPP2P323PPP2PPP1PPP2P \\
21PPPPPPP12P23P1PPPPPP13P3P11 \\
}
\noindent
where \emph{P} is a position where an agent can be spawned, $1, \, 2, \, 3$ are berries that are initial set to red, green, and blue respectively. A visual depiction of this map is given in Fig.~\ref{fig:allelo_reset}. There are a total of $384$ positions where berries can be spawned. There are sixteen agents in the environment at each time (unless one is removed due to a zap), each corresponding to a unique policy.

\subsubsection{Action Space}
The agent has three additional actions, a re-planting red, green, and blue berry varieties. Technically, these actions are implemented by beams that shoot forward up to three grid cells and are blocked by the first berry they hit, as depicted in Fig.~\ref{fig:games_figure}. If they hit a differently colored berry, then it gets replanted with the chosen color berry variety. The replanting actions have a cooldown time and after being used cannot be used again within the next two steps.

\subsubsection{Transition Dynamics and Reward Function}
Our world map in this environment is a toroid so there is no notion of a boundary of the map and all agent moves (up, left, down, right) and rotations transition the agent to the desired grid cell (unless two agents attempt to enter the same grid cell, in which case the tie is randomly broken). When an agent steps over a berry, that berry is eaten and the agent will receive a reward of $2$ if that berry matches its taste preference and a reward of $1$ otherwise. Eight of the agents have a taste preference for red and eight have a taste preference for green. 

If $r, \, g, \, b$ denote the respective numbers of red, green, and blue sites at which berries can spawn, each site of a particular color will spawn a berry at each time-step with probability $0.0000025 * c$ where c is the number of berries of that color. However, a berry cannot be spawned more frequently than every $10$ steps and so if a berry has just been eaten, the probability of spawning a berry at that location is 0 until 10 steps have passed. Additionally, berries cannot grow underneath agents, so if an agent remains atop a berry patch no berry will spawn there while the agent stands there.

Agents are initially spawned in colored grey.
Agents that successfully change the color of a berry will acquire the new berry color. If those agents then eat a berry, they have some probability of reverting back to grey. If we define the monoculture fraction $m$ as $m = \text{max} \{\frac{r}{r + g + b}, , \frac{g}{r + g + b}, \, \frac{b}{r + g + b} \}$ then the probability of reverting back to grey is $1 - m$. Thus, as the monoculture gets high agents are grey less often. This allows agents to remain colored once they achieve high monoculture fraction, which solves a potential issue wherein monoculture fraction gets high, reducing opportunities to color berries, and agents are then mistakenly identified as free-riders. We observed that without this feature agents would learn to rapidly re-color berries to prevent misidentification as free-riders and added this feature to remove this behavior.

\subsection{Cleanup With Startup Problem}
\subsubsection{Initial Map}
\begin{figure}
    \centering
    \includegraphics[width=0.4\textwidth]{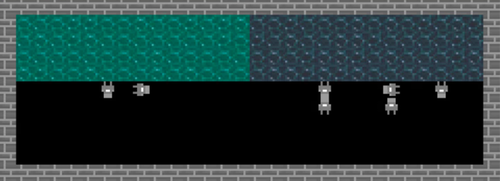}
    \caption{State of the map at initialization for Clean Up With Start Up Problem}
    \label{fig:cleanup_reset_screen}
\end{figure}

An ASCII representation of the initial map is:

\noindent
\texttt{WWWWWWWWWWWWWWWWWWWWWWWWWWWWWW \\
WFFFFFFFFFFFFFFDDDDDDDDDDDDDDW \\
WFFFFFFFFFFFFFFDDDDDDDDDDDDDDW \\
WFFFFFFFFFFFFFFDDDDDDDDDDDDDDW \\
WFFFFFFFFFFFFFFDDDDDDDDDDDDDDW \\
WPPPPPPPPPPPPPPPPPPPPPPPPPPPPW \\
WPPPPPPPPPPPPPPPPPPPPPPPPPPPPW \\
WBBBBBBBBBBBBBBBBBBBBBBBBBBBBW \\
WBBBBBBBBBBBBBBBBBBBBBBBBBBBBW \\
WBBBBBBBBBBBBBBBBBBBBBBBBBBBBW \\
WWWWWWWWWWWWWWWWWWWWWWWWWWWWWW}
\noindent

where P is a site where agents can initially be spawned on, B is a site where apples can spawn, W is a wall that cannot be passed, F is a cell of pollution type 1 and D is a cell of pollution type 2. An image-based depiction of this is in Fig.~\ref{fig:cleanup_reset_screen}.

\subsubsection{Action Space}
The agent has one additional action, a cleaning action. This beam shoots forward up to three grid cells and is blocked by the first bit of pollution it hits. The beam has a cooldown time of 5 seconds. Any cleaned tile has a probability of $0.05$ of spawning pollution on it. However, a tile that has been cleaned will not spawn pollution for the first 100 seconds after being cleaned.

\subsubsection{Transition Dynamics and Reward Function}

As in allelopathic harvest, we wanted to make it possible for agents to eat a few apples without being immediately labelled as a free-rider. Here we add a setting wherein an agent can eat up to 20 berries before being turned gray. If an agent eats 21 berries in a row without cleaning in between, it will turn back to grey.

Now we define the probability function that determines the rates at which apples regrow. If $d_1$ is the number of uncleaned tiles of color 1 and $d_2$ is the number of uncleaned tiles of color 2 then the probability of an apple spawning at any particular site is:
\begin{equation*}
    0.05 * \left(1 - 0.25 \left(\frac{d_1}{d_1 + d_2} *\frac{d_2}{d_1 + d_2}\right)^{10} \right)
\end{equation*}
This product between the two ratios creates the startup-up problem: if agents equally clear both types of pollution then they are no better off than if they had not cleared any pollution at all. Consuming an apple gives an agent a reward of $1$.

\section{Learning Dynamics}
\begin{figure}
\centering
\begin{subfigure}{.3\textwidth}
  \centering
  \includegraphics[width=\linewidth]{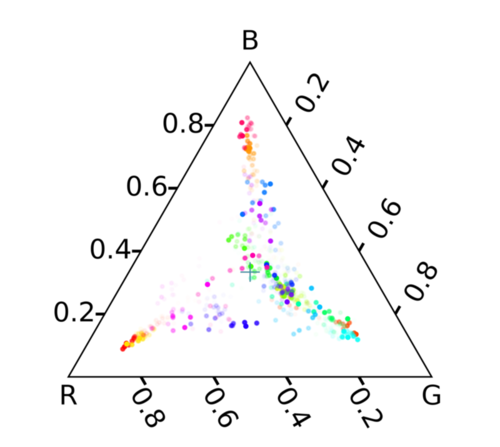}  
  \caption{}
  \label{fig:simplex_a}
\end{subfigure}
\begin{subfigure}{.3\textwidth}
  \centering
  \includegraphics[width=\linewidth]{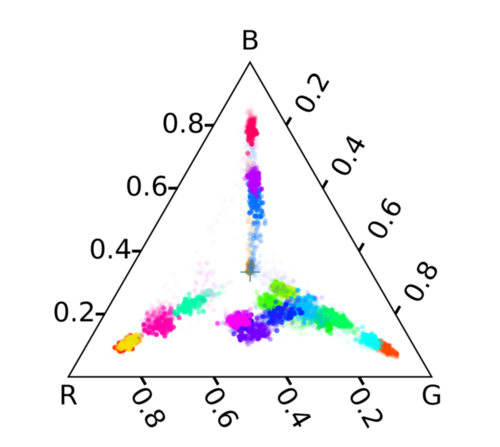}  
  \caption{}
  \label{fig:simplex_b}
\end{subfigure}

\begin{subfigure}{.3\textwidth}
  \centering
  \includegraphics[width=\linewidth]{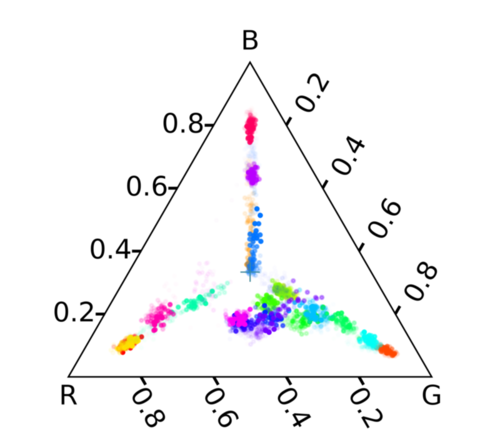}  
  \caption{}
  \label{fig:simplex_c}
\end{subfigure}
\begin{subfigure}{.3\textwidth}
  \centering
  \includegraphics[width=\linewidth]{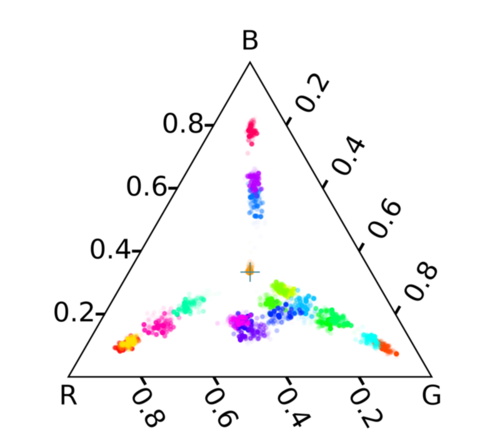}  
  \caption{}
  \label{fig:simplex_d}
\end{subfigure}
    \caption{Evidence of early learning and subsequent semi-stable planting behavior. The cross in the center represents equal berry fractions.  Individual dots are samples over a run where darker dots represent later points. (a) First 0 to $5e7$ time-steps. (b) $5e7$ to $1e8$ steps. (c) $1e8$ to $3e8$ steps. (d) $3e8$ to $5e8$ steps.}
    \label{fig:allelo_simplex_supp}
\end{figure}

This section demonstrates that in Allelopathic Harvest, the majority of the learning is confined to the first $1e8$ time-steps with only small changes occurring after that in a few seeds. Fig.~\ref{fig:allelo_simplex_supp} demonstrates that the majority of the change in planting behavior occurs early and the later time-steps only change in that 1 seed falls back to full free-riding and a few seeds slightly shift their planting position. 

\section{Computation Resources and Hyperparameter Tuning}

\subsection{Tuning Procedure}
Here we outline in brief the process by which we arrived at our final set of hyperparameters. The intent of this section is to provide the reader with a sense of the level of tuning that preceded any final hyperparameter selection. For both A3C and the contrastive predicting coding (CPC) unit we did not perform any tuning of their hyperparameters and only tuned hyperparameters of our classifier and pseudorewards.

\subsubsection{Tuning the classifier freeze}
The number of actor steps above which training the classifier was frozen was tuned by performing a run without any freezing and observing the point at which the balanced accuracy went above 0.9. For Allelopathic Harvest we only tested a cliff at $1e8$ while for Cleanup With Startup Problem (\emph{CSP}) we tested cliffs at $0.5e8$ and $1e8$ before settling on $0.5e8$.

\subsection{Tuning the size of the pseudorewards}
Although for the final experiments we run a fixed size of pseudoreward for each of the environments, there was a heuristic tuning period where we tested a few different hyperparamter magnitudes for each of the environments. Let $\alpha$ refer to the reward for punishing in accord with the classifier and $\beta < 0$ be the penalty for punishing in disaccord with the classifier. Then, for \emph{AH} we tested early on $\left(\alpha, \beta \right) \in \{ (0.2, 0.4), (0.4, 0.8), (0.8, 1.6) \}$ and for \emph{CSP} we tested  $\left(\alpha, \beta \right) \in \{ (1.0, 2.0), (1.2, 2.4) \}$. 

\subsection{Final Hyperparameters}
\label{sec:hyperparams}
For both \emph{AH} and \emph{CSP} we used the following shared hyper-parameters given in Table.~\ref{tab:common-parameters} where $\left[a, b \right]$ indicates that for a given seed the initial values from the hyperparamter will be drawn from a log-uniform distribution with probability density function $f(x; a, b) = \frac{1}{x \left[ \ln(b) - \ln(a)\right]}$

\begin{table*}[bt]
\centering
\begin{tabular}{cc}
\toprule
common hyperparameters      & value  \\
\midrule
learning rate & $\left[1e-4, 1e-3 \right]$ \\
entropy bonus & $\left[1e-3, 1e-2 \right]$ \\
batch size & 16 \\
$\gamma$ i.e. discount & $0.99$ \\
number of CPC steps & $20$ \\
CPC latent space dimension $l$& $64$ \\
CPC loss scaling $\lambda_C$ & $10.0$ \\
number of CPC steps (S) & $20$ \\
critic loss scaling $\lambda_{\text{critic}}$ & 0.5 \\
RMSProp $\epsilon $ & $1e-5$ \\
RMSProp momentum & 0.0 \\
RMSProp decay & $0.99$ \\
classifier loss scaling $\lambda_\text{class}$ & $0.01$ \\
classifier positive batch size & $32$ \\
classifier negative batch size & $1024$ \\

\bottomrule
\end{tabular}
\caption{Common hyperparameters used in Allelopathic Harvest and Cleanup With Startup Problem.}
\label{tab:common-parameters}
\end{table*}

\begin{table*}[bt]
\centering
\begin{tabular}{cc}
\toprule
common hyperparameters      & value  \\
\midrule
$\alpha$ & 0.2 \\
$\beta$ & 0.4 \\
Freeze step & $1e8$ \\

\bottomrule
\end{tabular}
\caption{Specific Hyperparameters used in Allelopathic Harvest}
\label{tab:allelo-parameters}
\end{table*}

\begin{table*}[bt]
\centering
\begin{tabular}{cc}
\toprule
common hyperparameters      & value  \\
\midrule
$\alpha$ & 1.0 \\
$\beta$ & 2.0 \\
Freeze step & $0.5e8$ \\

\bottomrule
\end{tabular}
\caption{Specific Hyperparameters used in Clean Up With Start Up Problem}
\label{tab:cleanup-parameters}
\end{table*}

The hyperparameters for Allelopathic Harvest are given in Table~\ref{tab:allelo-parameters} and for Clean Up With Start Up Problem in Table~\ref{tab:cleanup-parameters}.

Below we outline in more detail what each of the above terms mean. 
\begin{itemize}
    \item Freeze step: After this many environment steps, the classifier learning rate is set to 0.
    \item $\alpha$: the additional reward received when zapping an agent successfully when the classifier predicted a zap.
    \item $\beta$: the penalty received when zapping an agent successfully when the classifier predicted not to zap.
\end{itemize}

Finally, the loss is forming by combining the sum of the classifier loss, A3C loss, and CPC loss weighted by the loss scalings indicated in Table.~\ref{tab:common-parameters}.

\end{document}